# What is the physical meaning of mass in view of wave-particle duality? A proposed model

Donald C. Chang

*Macro-Science Group, Division of LIFS, Hong Kong University of Science and Technology, Clear Water Bay, Hong Kong, China*

Email: bochang@ust.hk

**ABSTRACT**

Mass is an important concept in classical mechanics, which regards a particle as a corpuscular object. But according to wave-particle duality, we know a free particle can behave like a wave. Is there a wave property that corresponds to the mass of a particle? This is an interesting question that has not been extensively explored before. We suggest that this problem can be approached by treating the mass on the same footing as energy and momentum. Here we propose that, all particles are excitation waves of the vacuum and different particles are represented by different excitation modes. Based on such a model, we found that mass is not an intrinsic property of the particle. Instead, mass is basically a measure of the particle energy. The relations between energy and mass can be directly derived based on the wave properties of the particle. This work explains why some particles are "wave-like" while others are "particle-like". Furthermore, this work has several interesting implications. It directly explains why photons can interact with a gravitational field. It also suggests a possible origin of dark matter; they are thought to be composed of excitation waves that fail to interact with each other. From this model, one can easily see why our universe has more dark matters than visible matters.

## 1. Introduction

It is well established in quantum mechanics that a particle (such as a photon or an electron) can behave like a corpuscular object as well as a wave. This peculiar property is called "wave-particle duality". Thus, an elementary particle does not always behave like a classical particle, which is regarded as a massive pointed object (like a bullet). Instead, an elementary particle sometimes can behave like a wave packet (i.e., like a light wave). Then, there is an important question: *Do the particle properties and the wave properties have a one-to-one correspondence?* In the cases of energy and momentum, there is no doubt that such one-to-one correspondence holds. For example, according to the Planck's relation $E = \hbar\omega$ and the de Broglie relation $\boldsymbol{p} = \hbar\boldsymbol{k}$, the particle properties $E$ (energy) and $\boldsymbol{p}$ (momentum) are corresponding to the wave properties $\omega$ (frequency) and $\boldsymbol{k}$ (wave vector). But what about mass? We know a particle has mass; could a wave have mass too? What is the physical meaning of mass in the wave view?

In fact, even in the particle view, the physical meaning of mass has not been very clear. Max Jammer, a leading historian of the concept of mass, wrote in 1999 that [1] "*. . . in spite of all the strenuous efforts of physicists and philosophers, the notion of mass, although fundamental to physics, is . . . still shrouded in mystery.*" Jammer was not alone in this opinion. According to a review recently written by John Roche entitled "*What is mass?*", there are still difficulties today with the understanding of the concept of mass [2]. In the conventional thinking of physics, mass is regarded as an intrinsic property of the particle. But then, how can mass vary with the particle's speed? And how can one explain why mass can be converted into energy and vice versa under certain situations?

One may say that these peculiar properties of mass are due to the effect of relativity. However, the special theory of relativity (STR) does not directly address the physical meaning of mass; it only deals with the changes of mechanical properties in different inertial frames. Furthermore, STR does not touch on issues related to the wave-particle duality.

In this work, we will investigate the possible physical meaning of mass based on a matter wave model. We think that mass can be understood on the same footing as energy and momentum. Using this model, one can easily see why the motion of a particle can deviate from Newtonian mechanics; it is a consequence of the fact that the elementary particle is essentially a wave packet in nature. In the followings, we will offer a detailed explanation of this view and discuss what interesting implications it may have.

## 2. The physical meaning of mass in the particle view

Before we discuss the physical meaning of mass in the wave view, let us first briefly review what is the physical meaning of mass in the traditional particle view.

### 2.1. The physical meaning of mass according to classical mechanics

The modern concept of mass probably started from Isaac Newton [3]. Before Newton, people only had a rough idea about mass based on its weight [4]. But they did not know what the physical basis of weight is. Newton made two major contributions in defining the physical concept of mass:

(a) He proposed a concept of *inertial mass*: such a mass plays a key role in his equation of motion:
$$F = ma \qquad (1)$$

(b) Furthermore, Newton invented the concept of *gravitational mass.* Based on Kepler's astronomical studies, Newton proposed a gravitational law
$$F = G\frac{m_1 m_2}{r^2} \;, \qquad (2)$$

in which $m_1$ and $m_2$ are the "masses" of two attracting bodies.

Thus, mass actually has two meanings in Newtonian mechanics: "inertial mass" and "gravitational mass". But according to Newton, the gravitational mass and the inertial mass were assumed to be the same. From then on, mass is regarded as an intrinsic property of a physical object. This becomes an important part of classical mechanics.

### 2.2. Einstein's view on the physical nature of mass

Einstein was the first one to realize that the mass of an object does not need to be a constant. It can vary with the traveling speed. In his famous 1905 paper on relativity [5,6], Einstein proposed that the mass of a traveling object can vary in the following way:

$$\begin{cases} \text{Longitudinal mass} = \dfrac{m}{\left(1 - v^2/c^2\right)^{3/2}} & (3) \\ \\ \text{Transverse mass} = \dfrac{m}{1 - v^2/c^2}. & (4) \end{cases}$$

In order to obtain the above equations, Einstein explicitly stated that his definition of mass was relying on Newton's law, $F = ma$. So, at least for the earlier days, Einstein's concept of mass did not deviate much from that of Newton.

### 2.3. Modern definition of mass in the particle view

Starting from the beginning of 20$^{th}$ century, physicists realized that concepts used in classical mechanics may not be totally applicable to all situations. For example, for events happening in a very small time and spatial scales, the physical laws would follow quantum mechanics instead of Newtonian mechanics. Thus, one needs to reconsider the definition of mass.

Nowadays, the concept of mass is mainly defined by relating it to momentum, that is,

$$p = mv. \qquad (5)$$

Thus, mass has one clear meaning: $m$ is the proportional constant between $p$ and $v$. One may say that this definition is an extension of Newton's concept of inertial mass. In modern days, the second law of Newton's equation of motion has been generalized to: Force = time derivative of momentum, i.e.,

$$F = \frac{dp}{dt}. \qquad (6)$$

Since $p = mv$,

$$F = m\frac{dv}{dt} + \frac{dm}{dt}v.$$

When $v$ is very small in comparison to the speed of light, $m$ can be taken as a constant. Under this situation, we have

$$F = m\frac{dv}{dt} = ma.$$

Thus, Newton's second law is only an approximation when the change of *m* is negligible.

## 3. What is the meaning of mass in the wave view?

As we stated earlier, one great discovery in modern physics is that an elementary particle can sometimes behave like a wave. In the case of photon, there is no doubt that it has all the wave properties, since it is a light wave. But even for particles with rest mass, such as electrons or neutrons, they also demonstrated wave properties. It is well known that electrons and neutrons can be diffracted from a crystal following Bragg's diffraction law [7,8]. They must behave like a wave. But then, how can a wave have mass?

This question can be answered easily if one accepts the definition of mass as outlined in Eq. (5), i.e., "mass" is only the proportional constant between *p* and *v*. We know that a wave has a momentum which is proportional to the wave number *k*:

$$p = \hbar k, \qquad (7)$$

where $k = 2\pi / \lambda$. ($\lambda$ is the wavelength). We also know that the energy of a wave is proportional to its frequency, i.e.,

$$E = \hbar \omega. \qquad (8)$$

Since the particle is a wave packet, its group velocity is given as

$$v = \frac{d\omega}{dk} = \frac{dE}{dp}. \qquad (9)$$

We can calculate *v* easily once we know the dispersion relation of a wave packet. Substituting Eqs. (7) and (9) into Eq. (5), we have

$$m = \frac{p}{v} = \hbar k \bigg/ \frac{d\omega}{dk} = p \bigg/ \frac{dE}{dp}. \qquad (10)$$

Thus, the mass can be calculated explicitly. This explains why a wave packet can have mass.

From the above equation, it is easy to see that mass is not necessarily a constant. Since the dispersion relation $\omega = \omega(k)$ is not linear in general, *m* is a function of *k* according to Eq. (10). That means the mass should vary with momentum change. Thus, it is not a mystery that the mass of a particle would change with speed. Instead, it is a natural expectation.

## 4. Mass of a photon

Let us examine first the physical nature of mass in a photon. For a long time, there has been an active debate on the physical property of light. On one hand, we know from the Maxwell's theory that light is a kind of electro-magnetic radiation. On the other hand, from the Planck's law in black-body radiation and Einstein's theory on photo-electric effect, we know light can behave very much like a particle. Then, does this particle have mass? If yes, what is the mass of a photon?

From Eq. (10), it is easy to see that, the key to determine the physical behavior of mass of a free particle is the dispersion relation between $\omega$ and $\mathbf{k}$ (i.e., between $E$ and $\mathbf{p}$). In the case of a photon, this dispersion relation is already known, that is,

$$\omega = 2\pi\nu = \frac{2\pi c}{\lambda} = ck \,. \tag{11}$$

By multiplying the left hand side and the right hand side of the above equation with $\hbar$, and using the Planck's relation and the de Broglie relation, Eq. (11) becomes

$$E = cp \,. \tag{12}$$

Substituting this into Eq. (10), we have

$$m = \frac{E/c}{c} = \frac{E}{c^2},$$

i.e.,

$$E = mc^2 \,. \tag{13}$$

Thus, it is apparent that the radiation wave (photon) can have a mass, and this mass is directly related to its energy.

One should notice that the mass defined in Eq. (13) is not the "rest mass" ($m_0$) of a particle; $m$ may be called the "moving mass". It is well known that a photon does not have rest mass. Other free particles, such as electrons, are known to have rest mass. Recent experiments suggested that neutrinos can also have nonzero rest mass[9]. What will be the physical meaning of mass for those free particles?

## 5. Mass of a free particle with rest mass

In order to investigate the physical origin of the rest mass of a free particle, we can again make use of Eq. (10). That means we need to find the dispersion relation of the matter wave which represents a particle with nonzero rest mass (such as an electron). But in order to do that, we must first derive the wave equation that describes the motion of the free particle in the vacuum. This is what we are about to do in the following sections.

### 5.1. The basic hypothesis of the matter wave model

From the early days of the 20th century, it was already known that sub-atomic particles can behave like waves. There is strong evidence suggesting a close similarity between the physical properties of an electron and that of a photon. For example, it was shown in 1927 that electrons can be diffracted by a nickel crystal following the Bragg's law [7]. This behavior is similar to photons. In fact, many pioneers in modern physics, including Einstein and de Broglie had proposed that the matter wave of a sub-atomic particle behaves very much like a photon. This thinking was also explicitly proclaimed by Richard Feynman in his famous *Lectures on Physics*: "……*electrons behave just like light. The quantum behavior of atomic objects (electrons, protons, neutrons photons and so on) is the same for all, they are all 'particle waves,'…so what we learn about the properties of electrons… will apply also to all 'particles,' including photons of light*". [10]

Thus, in this work, we hypothesize that:

> ***(1) Like the photon, a free particle (such as an electron) is an excitation wave of the vacuum.***
>
> ***(2) Different types of free particles are different excitation modes of the vacuum.***

One may notice that our proposal represents a slight modification from the traditional Copenhagen interpretation of the particle wave [11]. In most quantum mechanics textbooks, the particle is treated as a pointed object; the wave property is only associated with the probability of finding the particle at a particular space and time [12]. But in view of the findings of the diffraction experiments, it seems more sensible to regard the particle as a *physical wave* rather than a *probability wave*. The Copenhagen interpretation was based more on philosophical choices rather than physical evidence. Although such a statistical interpretation was strongly supported by some leaders in quantum mechanics, including Bohr and Heisenberg, it was not universally agreed. In fact, many well-known physicists, including Einstein, Schrödinger, and de Broglie, had opposed such an interpretation [11].

Here, we will simply hypothesize that, like the photon, the matter wave of a particle with non-zero rest mass is a real physical wave. Our hypothesis can be justified by a number of observations. Besides the fact that electrons can be diffracted from a crystal following the Bragg's diffraction law, the wave nature of electron is also clearly demonstrated in the operation of an electron microscope. Furthermore, we know particles can be created or annihilated in the vacuum. If the particle is a corpuscular object like a bullet, how can it be created from nowhere or disappear suddenly? The only possible explanation is that the particle is an excitation wave of the vacuum medium, so that it can be excited by an energetic stimulation and it can be transformed from one type of wave into another type of wave.

If one accepts our hypothesis, it is not difficult to see that the wave functions representing different kinds of particles are just different solutions of the basic wave equation describing the excitation of the vacuum medium. Then, our task is basically reduced to the following steps:

(1) To find a proper mathematical model to depict the physical properties of the vacuum
(2) To derive the wave equation that describes the motion of the excitation wave
(3) To solve this wave equation and obtain the dispersion relation

## 5.2. Physical nature of the vacuum

If the matter wave is a physical wave, then what is its carrying medium? In the 19[th] century, it was widely believed that the electro-magnetic field was carried by a medium called "aether" [13]. The aether hypothesis, however, was rejected in the early 20[th] century due to a number of difficulties [14]. To avoid any misunderstanding, we would like to emphasize that our concept of vacuum medium is different from the aether hypothesis. First, the aether was a hypothetical medium filling only the space between matters; the vacuum in our model is a pre-existing medium that fills all space in our universe. Second, the hypothetical aether is a medium for transmitting only EM radiation; the vacuum in our model is a medium for excitation waves representing all particles (radiation and matter). In another word, matters are composed of excitation waves of this vacuum medium. We have shown in an earlier work that the arguments against the aether hypothesis cannot be used to reject the existence of a vacuum medium [14].

The physical nature of the vacuum is a very complicated subject. At present, our knowledge about the vacuum is still limited. But based on the studies of electricity and magnetism, we know at least the long range behavior of the vacuum. According to the early work of Maxwell, the vacuum behaves basically like a dielectric medium [15,16]. In the original version of the Maxwell equations, the Ampere's Law was

$$\nabla \times \mathbf{H} = \mathbf{J} . \tag{14}$$

It had a problem, because it would lead to a violation of the condition of conservation of charge,

$$\nabla \cdot \mathbf{J} = -\frac{\partial \rho_e}{\partial t} . \tag{15}$$

In order to fix this problem, Maxwell later proposed to add a new term $\partial \mathbf{D}/\partial t$ into the right hand-side of Eq. (14) [16], i.e.,

$$\nabla \times \mathbf{H} = \mathbf{J} + \frac{\partial \mathbf{D}}{\partial t} . \tag{14A}$$

(**D** is called the "charge displacement".) The justification for this was that, when a dielectric medium is exposed to an electric field, it will induce a displacement of the dielectric charges. Such a displacement current ($\mathbf{J}_d = \frac{\partial \mathbf{D}}{\partial t}$) should be added into Eq. (14).

When Maxwell constructed his theory for light propagation in the vacuum, he assumed that the displacement current is the non-vanished current, i.e.,

$$\nabla \times \mathbf{H} = \frac{\partial \mathbf{D}}{\partial t} . \tag{14B}$$

This can be justified only if the vacuum is regarded as a dielectric medium!

The assumption of a displacement current was a stroke of genius. Without this assumption, it would be impossible to derive the wave equation for the propagation of electro-magnetic wave. Thus, in the Maxwell theory of radiation, it is essential to regard the vacuum as a dielectric medium.

### 5.3. The basic wave equation of a free particle

In the previous section, we showed that there is strong evidence indicating that electrons and photons behave similarly. Thus, we proposed that both the radiation waves and matter waves are excitation waves of the same vacuum medium. If this is the case, then waves representing different particles (with or without mass) should obey the same wave equation, which is derived based on the physical properties of the vacuum medium. At this point, we already know the wave equation of at least one particle; that is the photon. We know a photon is an oscillating wave of electro-magnetic fields, which are linear functions of the vector potential $A^\mu = (\phi, \mathbf{A})$. Using Maxwell equations, one can derive

$$\nabla^2 A^\mu - \frac{1}{c^2}\frac{\partial^2 A^\mu}{\partial t^2} = 0 \tag{16}$$

where $c = 1/\sqrt{\mu_0 \varepsilon_0}$ is the speed of light. ($\varepsilon_0$ is permittivity and $\mu_0$ is permeability in free space). This may provide a very useful hint. If matter waves and radiation waves are both excitation waves of the same vacuum medium, we expect that the wave function of the matter wave $\psi$ may also obey the same wave function as shown in Eq. (16), i.e.,

$$\nabla^2 \psi - \frac{1}{c^2}\frac{\partial^2 \psi}{\partial t^2} = 0. \tag{17}$$

We propose that this is the basic wave equation for all types of excitation waves of the vacuum. Different solutions of $\psi$ just represent different types of free particle in the physical world. The simplest solution of Eq. (17) is a plane wave $\psi_k \sim e^{i(\mathbf{k}\cdot\mathbf{x}-\omega t)}$, where $\mathbf{k}$ and $\omega$ are the wave vector and frequency, respectively. We propose that this plane wave solution represents a radiation wave (a photon) which is a particle without rest mass. This indeed is the common solution for the light wave equation as given in Eq. (16).

For a particle with non-zero rest mass, the solution is more complicated. Such a particle would behave like a mass point in the classical limit; thus, the probability of finding the particle should be highest at its trajectory. This suggests that the wave function of a massive particle not only depends on the spatial coordinate along its trajectory (i.e., $\hat{\mathbf{k}} \cdot \mathbf{x}$), it also varies in the transverse plane ($\hat{\mathbf{k}} \times \mathbf{x}$).

Thus, one may assume that the general wave function representing a free particle has the form

$$\psi_k(\boldsymbol{x},t) = \psi_T(\hat{\boldsymbol{k}} \times \boldsymbol{x})\psi_L(\hat{\boldsymbol{k}} \cdot \boldsymbol{x}, t), \tag{18}$$

where $\psi_L$ is the longitudinal component of the wave function which describes the travelling wave along the particle's trajectory, and $\psi_T$ is the transverse component of the wave function. Substituting Eq. (18) into Eq. (17), one can use the technique of separation of variables to obtain two coupled equations:

$$\begin{cases} \left[\nabla^2 - \dfrac{1}{c^2}\dfrac{\partial^2}{\partial t^2}\right]\psi_L(\hat{\boldsymbol{k}} \cdot \boldsymbol{x}, t) = \ell^2 \psi_L(\hat{\boldsymbol{k}} \cdot \boldsymbol{x}, t) & (19A) \\ \nabla^2 \psi_T(\hat{\boldsymbol{k}} \times \boldsymbol{x}) = -\ell^2 \psi_T(\hat{\boldsymbol{k}} \times \boldsymbol{x}), & (19B) \end{cases}$$

where $\ell$ is a fitting parameter which is subjected to the condition

$$\omega^2 = (k^2 + \ell^2)c^2. \tag{20}$$

To simplify the solution of Eqs. (19A) and (19B), let us denote the direction of $\boldsymbol{k}$ as the $z$-axis. By solving Eqs.(19A) and (19B) separately, one can show that the general solution of $\psi$ is

$$\psi_k(\boldsymbol{x}, t) = a J_n(\ell r) e^{\pm i n \theta} e^{i(kz - \omega t)}, \tag{21}$$

where $J_n$ is Bessel function of the first kind; $n$ is an integer or half integer; $r$ and $\theta$ represent the radial distance and the azimuthal angle of the space vector in the transverse plane. ($a$ is a normalizing constant.) As expected, the wave function of a free particle behaves like a travelling wave moving along the direction of its trajectory. But due to the presence of the Bessel function, $\psi_k$ varies in a diminishing oscillating manner in the directions perpendicular to the particle's trajectory.

### *5.4. Particle properties of the matter wave*

In order to verify that $\psi_k$ as described in Eq. (21) is a proper wave function representing the matter wave of a free particle, let us examine if it can produce the correct particle properties. Using the quantum correspondence rules $E \to i\hbar \partial/\partial t$ and $\boldsymbol{p} \to -i\hbar \nabla$, one can show that $\omega$ and $\boldsymbol{k}$ are related to the energy ($E$) and momentum ($\boldsymbol{p}$) of the particle, i.e.,

$$E = \int \psi_k^* i\hbar \frac{\partial}{\partial t} \psi_k \, d^3x = \hbar \omega, \tag{22}$$

and

$$\boldsymbol{p} = \int \psi_k^* \frac{\hbar}{i} \nabla \psi_k \, d^3x = \hbar \boldsymbol{k}. \tag{23}$$

(Here $\hbar$ is Planck's constant divided by $2\pi$). This reaffirms that, like the photon, a free particle with rest mass also obeys the Planck's relation and the de Broglie relation!

From the wave parameters in $\psi_k$, we can also obtain the particle mass. Eq. (20) can be rewritten as

$$(\hbar\omega)^2 = c^2\left(\hbar^2 k^2 + \hbar^2 \ell^2\right). \tag{20A}$$

Using the Planck's relation and the de Broglie relation, the above equation becomes

$$E^2 = c^2 p^2 + c^2 \hbar^2 \ell^2, \tag{24}$$

or,

$$E = \sqrt{c^2 p^2 + c^2 \hbar^2 \ell^2}. \tag{24A}$$

Recall that the particle velocity ($v$) is determined by the group velocity of the wave packet [12], $v = \dfrac{\partial \omega}{\partial k} = \dfrac{\partial E}{\partial p}$, and the particle mass $m$ is defined by $p = mv$ in the classical limit, one can use these relations to solve Eq. (24A) and obtain

$$m = \dfrac{\hbar \ell / c}{\left(1 - v^2/c^2\right)^{1/2}}. \tag{25}$$

We know at $v = 0$, $m$ equals the rest mass $m_0$. Eq. (25) then becomes

$$m_0 = \dfrac{\hbar \ell}{c}. \tag{26}$$

Substituting this into Eq. (24), we can easily see that the dispersion relation for a free particle is

$$E^2 = p^2 c^2 + m_0^2 c^4. \tag{27}$$

Combine Eq. (25) and (26), we can also see that the mass of a particle is speed-dependent

$$m = \dfrac{m_0}{\sqrt{1 - v^2/c^2}}. \tag{28}$$

By combining Eqs. (5), (27) and (28), one can further obtain

$$E = mc^2. \tag{29}$$

Thus, *for a simple particle with nonzero rest mass, it obeys the same energy-mass conversion law as what we found in photon*. At this point, we can easily see that, the photon is no different from any other simple particles with nonzero rest mass. It just happens that the rest mass of the photon is zero!

From Eq. (27), one can see that, at $v \to 0$ (i.e., $p \to 0$), the resting energy of the particle becomes

$$E_0 = m_0 c^2. \tag{30}$$

Thus, the relation between energy and momentum for a simple particle can be written as

$$E^2 = p^2 c^2 + E_0^2. \tag{27A}$$

We propose that this relation is generally true for all free particles. It is derived solely on the assumption that an elementary particle is an excitation wave of the vacuum.

### 5.5. Both energy and mass are related to the bending curvature of the vacuum

From the above discussion, it is apparent that the resting energy $E_0$ is contributed solely from the rest mass $m_0$, which in turn is related to the parameter $\ell$ in the wave function. This provides a very useful hint about the physical nature of the rest mass. This in fact suggests that the rest mass could be related to a geometrical property of the vacuum. To understand this point, let us first review the physical nature of energy and momentum according to the view of wave-particle duality. From the Planck's relation and the de Broglie relation, it is easy to see that the energy and momentum of a free particle are related to the periodicity of oscillation of the vacuum medium. More specifically, $E = \hbar \omega$ is related to the periodicity of oscillation in the time dimension, and $p = \hbar k$ is related to the periodicity of oscillation in the spatial dimension along the direction of the trajectory. Our finding that $m_0 = \hbar \ell / c$ may suggest that the rest mass is associated with some sort of oscillation periodicity too. Indeed, from Eq. (21), we see that the transverse component of the free particle wave function is described by a Bessel function, the asymptotic form of which is

$$J_n(\ell r) \to \left(\frac{2}{\pi \ell r}\right)^{1/2} \cos\left(\ell r - \frac{2n+1}{4}\pi\right). \tag{31}$$

(The variation of Bessel function is shown in Figure 1). Thus, $\ell$ can be regarded as the "transverse wave number" of the free particle, i.e., it is the inverse of the wavelength in the transverse oscillation, $\ell = 2\pi / \lambda_T$. (See Figure 1). Thus, Eq. (26) means that the rest mass of a particle is associated with the oscillation periodicity of the wave function in the transverse plane.

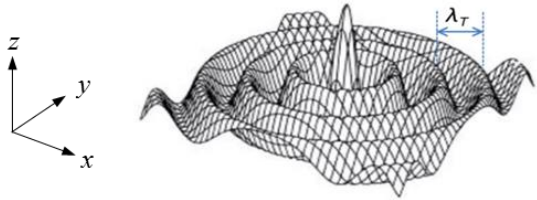

*Figure 1. A plot of the transverse component of the wave function $\psi_T$ for n = 0. The wavelength is denoted $\lambda_T$. The particle is traveling along the axis z.*

This result appears to make very good sense. In essence, our analysis suggests that *energy, momentum and mass are all related to the curvature of bending the vacuum medium during the propagation of the excitation wave*. Such bending curvatures are just taking place in different dimensions.

This result is highly interesting; it is also not very difficult to understand. Both energy and mass must be created by "work". When there is no excitation wave, there is no stress in the vacuum medium; its curvature is zero. When an excitation wave emerges, it creates a temporal stress at a local region of the vacuum medium. Since it takes work to bend the wave medium, the larger the bending curvature, more work is required. This is true for bending the medium in all dimensions (spatial and temporal). Thus, a shorter wavelength of a propagating wave should always associate with a higher "energy state", which may be reflected in an increase in energy, momentum or "mass" of the excitation wave.

### 5.6. The resting energy and the moving energy of a single particle appear to form a two-dimensional Hilbert space

The dispersion relation as given in Eq. (27A) provides an important insight about the physical nature of energy of a free particle. Namely, it suggests a special geometrical linkage between the particle energy and its momentum. For a moving particle, there are two types of energy associated with it: (1) The "moving energy" $E_K$, which associates clearly with the momentum, i.e., $E_K = cp$. (2) The "resting energy" $E_0$, which is the intrinsic energy of a particle that does not depend on its momentum. According to the Eq. (27A), the total energy of the particle ($E$) is not a linear combination (algebraic sum) of $E_0$ and $E_K$, instead, they form a triangular relationship following the Pythagoras law, i.e.,

$$E^2 = p^2c^2 + E_0^2 = E_K^2 + E_0^2. \qquad (27B)$$

This suggests that *the moving energy $E_K$ and the resting energy $E_0$ form a two-dimensional Hilbert space. The particle energy ($E$) is a vector sum of $E_K$ and $E_0$*. (See Figure 2).

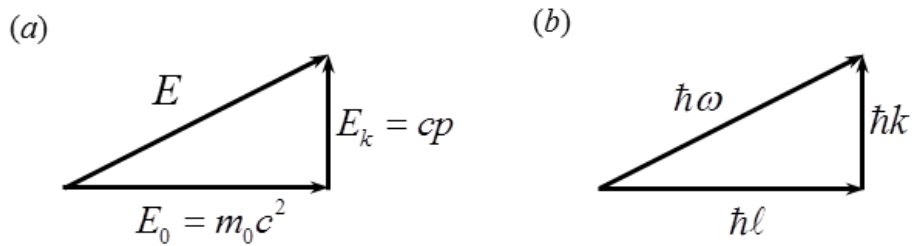

*Figure 2. The resting energy and moving energy of a single particle form a 2-dimensional Hilbert space. (a) Geometrical relationship between the corpuscular properties E, p and m. (b) Geometrical relationship between the wave parameters ω, k and ℓ. (Here we use the natural unit, c =1).*

This relationship can be understood easily from a wave perspective. Recall that Eq (27) was derived from the dispersion relation of a free particle, i.e., Eq (20A). Using the natural unit ($c=1$), Eq. (20A) can be rewritten as

$$(\hbar\omega)^2 = (\hbar k)^2 + (\hbar \ell)^2. \tag{20B}$$

It suggests that $\hbar\omega$ is a "vector sum" of two perpendicular vectors with amplitudes equal to $\hbar k$ and $\hbar \ell$. (See Fig. 2b). We know $\boldsymbol{k}$ is the wave vector parallel to the trajectory of the particle, while $\ell$ is a "wave number" that characterizes the oscillation of the wave function in a plane transverse to $\boldsymbol{k}$. Thus, the directions of oscillations in the vacuum characterized by $\boldsymbol{k}$ and $\ell$ apparently are associated with two perpendicular axes. We have demonstrated that $\hbar\ell$ is associated with the rest mass and thus the resting energy, and, from the de Broglie relation, we know that $\hbar k$ is associated with momentum and thus the moving energy. Hence, it is not surprising that the resting energy $E_0$ and moving energy $E_K$ may form a two-dimensional Hilbert space. Naturally, their vector sum becomes the total energy of the particle.

This provides a simple explanation to the physical nature of the mechanical properties of a particle. Furthermore, this geometrical interpretation suggests that the physical natures of energy and mass are very similar. There is no wonder why energy and mass can be converted between each other.

## 6. Mass of a composite particle

In our physical world, not all elementary particles are free particles; and not all free particles are elementary particles. For particles that we have discussed so far, like the electrons and photons, they are elementary particles and also they can travel in the space freely. Other elementary particles, such as quarks, cannot travel alone freely and can only be confined with other quarks to form a complex structure. They are not free particles. Some sub-atomic particles, such as protons and neutrons, on the other hand, can travel freely by themselves. But, they are not elementary particles. They are believed to be composed of more fundamental elementary particles, such as quarks. These particles can only be called "composite particles".

Thus, there are two different types of sub-atomic particles in our universe:

(1) <u>Simple particles</u>, like photons and electrons; that are stand-alone elementary particles with no internal sub-component. These simple particles also include other leptons, such as neutrinos and muons. These particles can move freely in space.

(2) <u>Composite particles</u>, like protons or neutrons; they are called "hadrons". Hadrons are not elementary particles, instead, they are composite particles made up of quarks. (Many other unstable particles generated in the particle-particle colliding experiments are also composite particles. For example, a positively charged pion is made up of two quarks: $u\bar{d}$.)

### 6.1. The energy-momentum relationship for a composite particle

In the last section, we discussed mainly the origin of mass in simple particles. For mass of a composite particle, it is much more complicated. We do not know the internal structure of a composite particle. It is well known that both long-range force (EM force) and short-range forces (strong interaction and weak interaction) are involved in the vacuum. We propose that the simple particles are excitation waves of the vacuum medium driven by the long-range force. Their wave equation is given in Eq. (17). The wave function we derived for photon and electron does not apply to quarks, neutrons or protons. These composite particles are supposed to be excitation waves driven by the short-range forces. In order to write down their wave equation, one needs to know how the weak interaction and strong interaction may affect the oscillation of the vacuum. We do not know that yet.

Thus, it is not possible to write down a simple wave equation for hadrons at present. But this may not prevent us to probe into their basic particle properties. Based on experimental observations and from general analogies between free particles (simple and composite), we can still obtain some basic information about the relationship between energy, momentum and mass for the composite particles. Like the simple particles, we know the composite particles also behave like waves. As we have mentioned before, electrons can be diffracted from a crystal following the Bragg diffraction law [7]. It was found later that neutrons and protons can also be diffracted in a similar manner as electrons [8,17]. This implies that composite particles also have wave properties like the electron and they follow the de Broglie relation too.

This similarly allows us to speculate that *the energy-momentum dispersion relation for a composite particle is similar to that of a simple particle*. In the last section, we already showed that, for a simple particle such as an electron, the resting energy and moving energy of a single particle appear to form a two-dimensional Hilbert space. The total energy of the particle is the vector sum of these two energies. We propose that the same is true for a composite particle. That is, Eq. (27A) is also applicable for the composite particles,

$$E^2 = c^2 p^2 + E_0^{\,2} \ . \tag{32}$$

Here, $E_0$ is the zero-point energy of the composite particle, which is the amount of energy possessed by the particle when there is no translational motion. What is the source of the resting energy? A natural expectation is that it must be contributed by the intrinsic energy of individual quarks and their binding energies. If the quarks inside the composite particle are not stationary, their motion could also contribute to a part of the resting energy.

### 6.2. Derivation of the relation between energy and mass for a composite particle

Once we know the relation between energy and momentum, we can derive the relationship between energy and mass. This can be done by differentiating Eq. (32) versus *p*,

$$2E\frac{dE}{dp} = 2c^2 p$$

Since the particle is a wave packet, $dE/dp = d\omega/dk = v$. Recall that $p = mv$, the above equation becomes

$$E = mc^2 . \tag{33}$$

At $v = 0$, $p = 0$, then

$$E_0 = m_0 c^2 . \tag{34}$$

Thus, no matter whether a free particle is a simple particle or a composite particle, one can always define a rest mass which is directly related to the resting energy. And, by substituting Eq. (34) into Eq. (32), we can obtain the "relativistic" energy-momentum relation for a composite particle,

$$E^2 = c^2 p^2 + m_0^2 c^4 . \tag{35}$$

Using Eq. (33) and $p = mv$, this can further give

$$m = \frac{m_0}{\sqrt{1 - v^2/c^2}} . \tag{36}$$

By comparing Eqs. (33 to 36) with Eqs. (27 to 30), it is clear that the relations between $E$, $p$ and $m$ are the same for both simple particles and composite particles.

### 6.3. The atomic nucleus can be regarded as a composite particle

Next, let us consider what the relations between $E$, $p$ and $m$ are for the atomic nucleus. In hydrogen atom, its nucleus is a proton. That means a hadron could be an atomic nucleus. This suggests that there can be some sort of similarity between a nucleus and a hadron.

In the traditional view, the atomic nucleus is thought as a cluster of nucleons (protons and neutrons). (See Figure 3a). But this may not be the true picture. Since nucleons are supposed to be held together by the force of strong interaction, and so are the quarks within an individual nucleon, there may not be a clear boundary between neighboring nucleons. Quarks belonging to different nucleons could interact just like quarks within the same nucleon. If this is the case, the entire atomic nucleus is just an aggregation of interacting quarks, instead of a cluster of protons and neutrons. (Fig. 3b).

Take the helium nucleus as an example. It is not necessary to be a composite of two individual protons and two neutrons. It can be a "cage" containing 12 quarks (6*u* and 6*d*); these quarks are moving rapidly inside it. In other words, the atomic nucleus is just an aggregation of entangled quarks; there is no individual proton or neutron inside it. The proton and neutron may appear as individual particle only when they are outside of the nucleus.

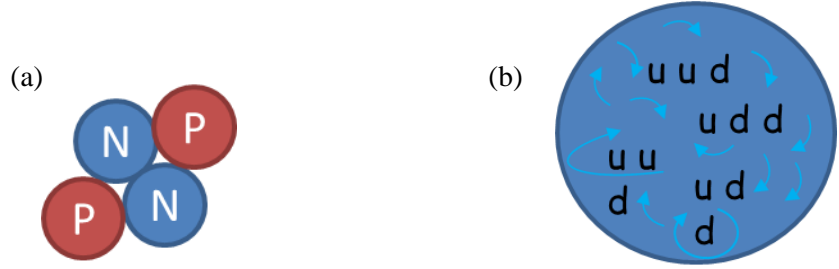

*Figure 3. Two views of the helium nucleus structure. (a) The traditional view of the helium nucleus. (b) We think in reality the helium nucleus is more like an aggregation of entangled quarks.*

From such a view, there is no conceptual difference between the nucleus of helium atom and the nucleus of hydrogen atom (i.e., a proton). They are all composite particles of quarks. This can also be true for nuclei of higher atomic numbers. This means that a nucleus can be generally regarded as a composite particle. Indeed, for radioactive isotopes emitting alpha particles, the radiation particle is identical to the nucleus of the helium atom.

Thus, we can expect that a nucleus under motion will have similar wave properties as a hadron. This expectation is indeed confirmed in experiments. Diffraction experiments using particle beams had indicated that, like protons and neutrons, helium atoms were also found to behave as waves and follow the de Broglie relation [8,17].

This suggests that Eqs. (32) to (36) are applicable also to the nucleus of an atom. The relationships between energy, momentum and mass are no different between an atomic nucleus and a hadron particle.

## 6.4. Common features for different types of particles in regarding to their relations between energy, momentum and mass

To summarize the above results, we conclude that, for particles of different kinds, including simple particles like photons and electrons, and composite particles like protons and atomic nuclei, they all obey the same energy-momentum relation, i.e.,

$$E^2 = c^2 p^2 + E_0^2 \ . \tag{32}$$

Furthermore, within the above equation, we also know

$$p = mv,$$

and

$$E_0 = m_0 c^2 . \tag{34}$$

Eq. (32) shows that $E^2$ and $p^2$ are in a linear relation. One can see that the energy-momentum relationships for different types of free particles are basically the same; they appear as parallel lines (see Fig. 4). The only difference between particles is that, the intersects of these lines with the vertical axis are not the same. That is, the resting energy $E_0$ for different particles is different. Since the resting energy is directly related to the rest mass by $E_0 = m_0 c^2$, this reflects that the rest mass is different for different kinds of free particle.

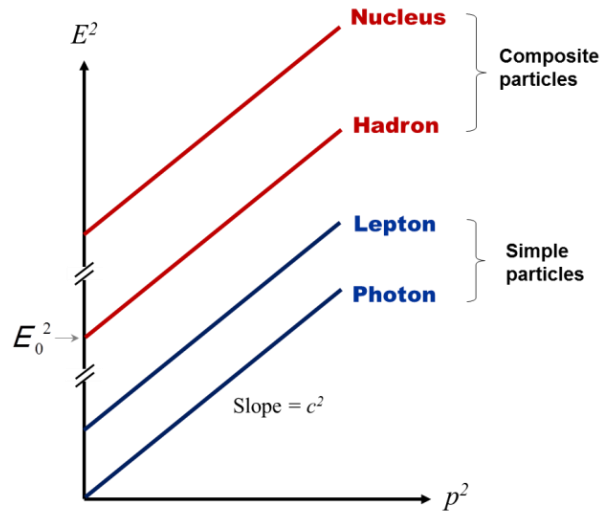

*Figure 4. Dispersion relations of energy and momentum for different particles.* *The slope is the same for all particles (slope = $c^2$), but the intercept $E_0^{\,2}$ is different for different particles. Since $E_0 = m_0 c^2$, this means that $m_0$ is different for different particles.*

From Figure 4, one can see that the slope of the energy-momentum relation is the same for all particles (slope = $c^2$), This suggests that both the simple particles and the composite particles are excitation waves of the same medium. We know the propagation speed of a wave is determined by the physical properties of the wave medium. Since the slope of photon is the same as that of electron (or hadrons), there is no basic difference between the radiation wave and the matter wave. They are just different excitation modes of the vacuum medium! Thus, the results plotted in Figure 4 are consistent with the basic assumption of our model that all free particles in our universe are excitation waves of the vacuum.

As demonstrated in the previous sections, the origin of the rest mass is different between a simple particle and a composite particle. For the simple particle (such as an electron or a neutrino), the rest mass is associated with a transverse wave number $\ell$. For the composite particles (such as a proton or neutron), the rest mass is apparently associated with the internal energy of the constituents which make up the composite particle. For example, a proton is thought to be made up of two *up* quarks and one *down* quark. Each of these quark is an elementary particle; it has its own resting energy (rest mass). But in addition to that, there are also binding energies between quarks. These energies are associated with the strong force needed to hold the quarks together. In fact, the resting energies of the quarks only account for a small fraction of the rest mass of the proton (See Table 1). The majority of the proton rest mass is thought to be contributed by the binding energies.

*Table 1. Mass of two hadrons and their containing quarks*

| **Particles** | **Mass** (MeV) | **Reference** |
|---|---|---|
| Proton (*uud*) | 938.27 | Ref.18, p.107 *(URL: http://pdg.lbl.gov)* |
| Neutron (*udd*) | 939.57 | Ref.19 |
| *Up* quark | 1.8-3.0 | Ref.18, p. 33 |
| *Down* quark | 4.5-5.3 | Ref.18, p. 33 |

Note: The mass listed here is actually the equivalent energy, i.e., $E_0 = m_0 c^2$.

## 7. The transition from wave-like to particle-like

From Figure 4, we can see that different particles have similar energy-momentum relations. Then, why some particles are wave-like while others are particle-like? We know some particles (such as photons or neutrinos) behave very much like waves; while other particles (such as protons, neutrons or atomic nuclei) behave more like massive pointed objects. What is the physical condition that determines whether a particle behaves like a wave or like a corpuscular object?

For a wave-like object, its energy-momentum relation will approach that of a photon, i.e., $E = cp$. For a particle-like object, its energy-momentum relation should approach what was described in the Newtonian mechanics. By carefully examine Eq. (32), we can easily see that the major factor determining whether a particle is wave-like or particle-like is essentially the balance between the moving energy (*cp*) and the resting energy ($E_0$). The following is a detailed analysis.

First, let us examine the case when $cp \gg E_0$. Eq. (27A) now becomes

$$E^2 = c^2 p^2 + E_0^2 \approx c^2 p^2. \tag{37}$$

Then,

$$E \approx cp. \tag{38}$$

This is similar to the energy-momentum relationship of a photon. Thus, the particle behaves very much like a wave.

Second, let us examine what happens when the moving energy of the particle is much smaller than the resting energy (i.e., $cp \ll E_0$). Under this situation, the energy-momentum relationship will approach that of Newtonian mechanics. Using the Taylor expansion, one can show from Eq. (27A),

$$E = E_0 \left(1 + \frac{c^2 p^2}{E_0^2}\right)^{1/2} = E_0 \left(1 + \frac{1}{2}\frac{c^2 p^2}{E_0^2} + ...\right) \approx E_0 + \frac{1}{2}\frac{c^2 p^2}{E_0}. \tag{39}$$

Since $E_0 = m_0 c^2$, we have

$$E = E_0 + \frac{1}{2}\frac{p^2}{m_0}. \tag{40}$$

In the classical limit, the *kinetic energy* is defined as $E - E_0$. Thus, the kinetic energy for a particle is

$$\text{kinetic energy} = \frac{1}{2}\frac{p^2}{m_0} = \frac{1}{2} m_0 v^2, \tag{41}$$

which agrees exactly with the relation in Newtonian mechanics. We can thus see that, whether the particle is wave-like or particle-like is mainly determined by a comparison between $E_0$ and $cp$. (See Figure 5). If $cp \gg E_0$, it is wave-like. On the other hand, if $cp \ll E_0$, it is particle-like!

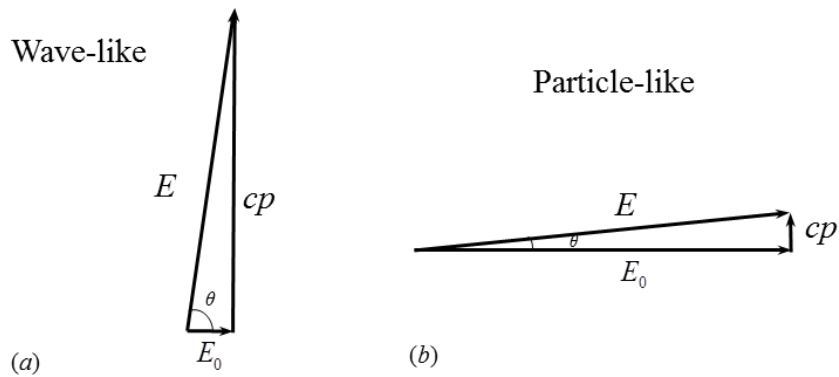

*Figure 5. Energy-momentum relationship for a free particle according to the matter wave model. The energy is the vector sum of cp and $E_0$. (a) In general, the energy-momentum relationship does not follow Newtonian mechanics. (b) When $cp \ll E_0$, the energy-momentum relationship will approach that of Newtonian mechanics; the particle now behaves more like a classical particle (i.e., a massive pointed object).*

To make things more simple, we can define a parameter called the "momentum-energy angle" ($\theta$), (see Figure 5),

$$\tan\theta = \frac{cp}{E_0}. \tag{42}$$

When $\theta$ is approaching $\pi/2$, the particle is "wave-like". On the other hand, if $\theta$ is approaching 0, the particle is very much "corpuscle-like". Its motion will follow the Newtonian mechanics. This explains that, for heavy particles, such as protons or atomic nuclei, they behave like a classical object. And for particles with very little or no rest mass, they behave like waves. For particles in between, they will behave partially like a wave and partially like a particle. This explains the physical basis for the observation of wave-particle duality.

## 8. Energy is a linear sum only in the classical limit

From the above discussion, we see there are two kinds of energy for a particle: the resting energy $E_0 = m_0 c^2$ and the moving energy $E_K = cp$. This moving energy is contributed solely by the momentum of the particle and thus one may think that it is equivalent to the kinetic energy in classical mechanics. However, this is not exactly the case, because there is a subtle difference between the two. For a free particle, the resting energy $E_0 = m_0 c^2$ and the moving energy $E_K = cp$ form a two-dimensional Hilbert space. The total energy of the particle ($E$) is a vector sum of $E_0$ and $E_K$, i.e.,

$$E^2 = E_K^2 + E_0^2. \tag{43}$$

Thus, it is not a linear relationship between $E$ and $E_K$. This is different from classical mechanics, in which the kinetic energy ($E_{KE}$) is defined as

$$\text{Kinetic energy } (E_{KE}) = E - E_0. \tag{44}$$

So, there is a linear relationship between the particle energy ($E$) and the kinetic energy in the classical limit. Furthermore, the explicit values of $E_K$ and $E_{KE}$ are not the same. For a free particle, the moving energy is defined as $E_K = cp$, while the kinetic energy of a particle in the classical limit is

$$E_{KE} = \frac{1}{2}\frac{p^2}{m_0}. \tag{45}$$

Only in the classical limit, when $cp \ll E_0$, the particle energy becomes a linear sum of the resting energy and the kinetic energy (see Eq. (40)).

Under the classical limit, the description of a physical system can become relatively simple. In our physical world, we know matters are made of atoms and molecules. Atoms, on the other hand, are made up of sub-atomic particles. We can roughly classify different physical objects based on their physical complexity:

(1) <u>Simple particles</u>, like electrons, which are stand-alone elementary particles.
(2) <u>Composite particles</u>, like neutron, proton and atomic nuclei, which are aggregates of elementary particles.
(3) <u>Physical systems</u>, like a tennis ball, a box or a spaceship, which are bulk objects made up of atoms.

So far, we have only talked about the energy of a free particle (simple or composite). But for most bulb objects, they are physical systems instead of particles. What are their relations between energy, momentum and mass?

We think these physical systems can be approximated as classical objects. That is because: (a) The moving speed of these objects is usually much slower than the speed of light. (b) These objects are made up of atoms; their nuclei have large masses and thus contain a significant amount of resting energy. They generally satisfy the condition $cp \ll E_0$. Thus, the energy-momentum relationship can be well described by the classical relation. This means that the entire object can be treated as a collection of mass points which obey the Newtonian mechanics.

Thus, the total energy of the bulk object can be regarded as a linear sum of the resting energy and kinetic energy of all constituent particles of that object. But in addition to that, there can be extra energy involved in the physical system. These may include various types of potential *energy $E_v$*, which arise from the interactions between particles within the system or due to the interactions between the object and external fields.

Thus, for a physical system, besides the sum of energy of all sub-atomic particles, its total energy should also include the potential energy of the system, i.e.,

*Total energy of the system = Total resting energy of particles*
*+ Total kinetic energy of particles*
*+ Total potential energy (between all particles)*
*+ Total potential energy (between particles and external fields)*

During a physical reaction, if the resting energy of all particles does not change, then the sum of their kinetic energies and potential energies will remain the same. This is known as *the law of conservation of energy*. From the discussion above, this law apparently applies only under the condition $cp \ll E_0$.

## 9. Conservation of energy and mass

### 9.1. Mass is a measure of energy instead of an independent physical entity

For a long time, mass has been thought as an intrinsic property of matter. But from the above discussion, one can see that this picture does not seem to be correct. Our analysis suggests that mass should be treated on the same footing as energy and momentum. As we have pointed out in Section 6, the rest mass of the atomic nucleus is not an "independent" physical entity. Instead, it is just a measure of the amount of resting energy associated with the nucleus, i.e.,

$$m_0 = \frac{E_0}{c^2}. \tag{46}$$

The resting energy $E_0$ here is sum of multiple contributions, which include the resting energies of individual constituents (quarks) and their interacting energies. The rest mass $m_0$ does not have an independent physical meaning. It is called "mass" simply because it satisfies the momentum-speed relationship as defined in Eq. (5), i.e., $p = mv$. This can be demonstrated very easily. Recall that because the particle is a wave packet, its speed is $v = \frac{\partial \omega}{\partial k} = \frac{\partial E}{\partial p}$. From

$$E^2 = c^2 p^2 + E_0^2, \tag{32}$$

we can differentiate both sides with respect to $p$ and obtain

$$v = \frac{\partial E}{\partial p} = \frac{c^2 p}{E}.$$

At low speed ($v \ll c$), $E \to E_0$, the above equation becomes

$$p = \frac{E_0}{c^2} v = m_0 v. \tag{47}$$

This suggests that $E_0 / c^2$ is playing the role as a "*rest mass*". So, $m_0$ does not have an independent origin, it is just a measure of the resting energy $E_0$.

Furthermore, we should notice that, the mass defined in Eq. (5) is not a constant. This is another indication that mass is not an intrinsic property of matter. Figure 6 shows the general relationship between $p$ and $v$. The mass $m$ is defined as the *slope* of the line connecting the origin with the $p$ versus $v$ curve. It is apparent that this slope can vary with $v$. Then, for a particle under acceleration, its mass is not a constant. If the mass $m$ is an intrinsic property of an object, its value should not change during motion.

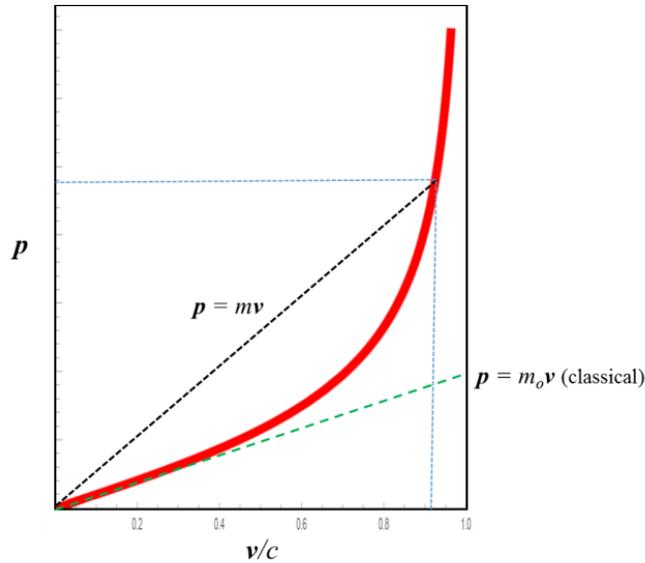

*Figure 6. The relationship between momentum and velocity according to the matter wave model.* Here, the red line is a plot of the variation of p as a function of v according to $p = m_0 v (1-v^2/c^2)^{-1/2}$. The mass m is defined by the slope of the relation $p = mv$ (black dashed line). In classical mechanics, the momentum is assumed to be $p = m_0 v$ (green dashed line).

In the older days, people did not know that *m* can vary with speed. They had the impression that *the mass is an intrinsic physical entity.* But such an impression is just an illusion.

From the above discussion, it is clear that the particle view used in classical mechanics is not an accurate description of the dynamic process in the microscopic scale. This is because a particle is actually a wave. Its energy, momentum and mass relationships in general are different from that of Newtonian mechanics.

### *9.2. The rest mass is not conserved during particle-particle interactions*

Because the rest mass is just a measure of the resting energy of a particle, it is not necessarily conserved during a particle-particle interaction. For example, when an electron and positron collide with each other to create a photon, the rest mass of electron and positron do not transfer to become the rest mass of the photon. Similarly, when an electron-positron pair is created, their resting energy did not come from the photon. This is because *only the total energy before and after the reaction is conserved, the resting energies of the participating particles are not conserved*. Hence, the total rest mass before and after the annihilation/create process can be different.

A change of the resting energy generally happens in most nuclear reactions. For example, many isotopes are radioactive. These isotopes can emit particles (e.g., *α, β, γ*) from the nuclei. During such a radioactive activity, the resting energy of the nucleus would change. That means the rest mass of the nucleus is no longer the same as before.

In some cases, the difference in resting energies before and after the nuclear reaction could be significant. Take the example of the fission of a uranium atom,

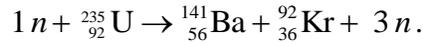

$$1n + {}^{235}_{92}U \rightarrow {}^{141}_{56}Ba + {}^{92}_{36}Kr + 3n.$$

The sum of the resting energies of the fission products is significantly less than the resting energies of the neutron and the uranium atom. Thus, about 200 MeV of energy will be released during this nuclear reaction [20]. Such energy could be used to generate electricity (or to make a nuclear bomb).

In summary, during a particle-particle interaction, only the total energy is conserved; the resting energy is not conserved. The resting energy of the participating particles can be converted into moving energy and vice versa.

## 10. Interpretation of Newton's gravitational law from the wave view

As we have discussed earlier, *mass* can have two meanings in classical mechanics, one is *inertial mass* and the other one is *gravitational mass*. In the above discussion, we showed that the concept of *inertial mass* could be an illusion. This so called "*inertial mass*" is actually a measure of the energy of a particle, i.e., $m = E/c^2$. This explains why a particle made up of wave can have mass.

Now, if one accepts that mass is not an intrinsic property of the particle but a measure of the particle energy, how can one explain that mass is the source of the gravitational force? In Newton's gravitational law, the gravitational force between two objects is proportional to the product of their masses. If mass is no longer regarded as an intrinsic property of the particle in the wave view, how can one explain the generation of the gravitational force?

I think the answer could be: Newton's gravitational law can be rewritten based on the energy of the object rather than the mass of the object. Since the mass *m* is proportional to the energy of a particle *E* by the relation $E = mc^2$, we can express Newton's gravitational law as

$$F = G\frac{m_1 m_2}{r^2} = \frac{G}{c^4}\frac{E_1 E_2}{r^2} = G'\frac{E_1 E_2}{r^2}, \qquad (48)$$

where $G' = G/c^4$. In another word, we can interpret the source of the gravitational force as *energy attracting energy* instead of *mass attracting mass*!

This energy here is the total energy of the object, which includes both the resting energy $E_0$ and the moving energy *cp*. Thus, for particles that have no resting energy, they can still be attracted in a gravitation field. This explains why light can be bent when it pass through the vicinity of a massive object. This has been verified in astronomical observations, as we know there is a "lensing effect" when light passes through a galaxy [21-23].

## 11. Discussion

### *11.1. Mass should be treated on the same footing as energy and momentum*

What is the meaning of mass? This is an important question in the history of physics [1]. Newton is probably the first one to give a scientific concept of mass. In his 1687 work "*Mathematical Principles of Natural Philosophy*" [3], he thought "mass" is "the quantity of matter". He found that for any two objects, the ratio for their inertia and the ratio for their weight are the same. This implies that the inertia mass and the mass associated with weight are equal. Then, people could measure the mass of a body by determining its weight.

Furthermore, Newton proposed that the weight of an object is just a measure of the gravitational force for that object. This then implies that the inertia mass and the gravitational mass are the same thing.

Today, one can measure mass using several distinct phenomena. This makes some theorists to speculate that mass could have different meanings:

- **Inertial mass** (which measures an object's resistance to being accelerated by a force)
- **Active gravitational mass** (which measures the gravitational force exerted by an object)
- **Passive gravitational mass** (which measures the gravitational force exerted on an object in a known gravitational field)
- **Energy-mass** (which measures the total amount of energy contained within a body)

In this work, we think there is only one "*mass*"; it is *a measure of the energy of the particle*. Such an interpretation is consistent with Einstein's view that mass can be interpreted as "a reservoir of energy" [24]. We show here that, in the classical limit, such an energy-related mass will automatically appear in the equation of motion as an *inertial mass*. Also, in Section 10 of this work, we showed that the gravitation law can be interpreted as *energy-attracting-energy*; it only gives an appearance of *mass-attracting-mass*. The perception of "gravitational mass" can be regarded as an illusion.

In Newtonian mechanics, the mass is regarded as an intrinsic property of an object. Why? From human observation, it is very easy to see that any object is made up of "matters". Such matter could be a piece of rock, a sheet of metal, or a piece of wood. Each of these matters has a certain weight. From a long time ago, people thought that such weight must come from the "mass" of the object. An object with more mass would have more weight. Thus, it is very natural for ancient people to think that the mass is an intrinsic property of an object.

In this work, we propose that a particle is an excitation wave of the vacuum, and, particle properties including energy, momentum and mass can all be treated on the same footing. This proposal is mainly based on the phenomenon of wave-particle duality, namely, a particle has both wave and corpuscular properties. We suggest that, like the energy and momentum, mass has a corresponding meaning in the wave view. We show that the equation of motion in classical mechanics is only an approximation under the condition that the "moving energy" of a particle is much smaller than its "resting energy".

Based on this wave model, we think mass is not more intrinsic than energy and momentum. The main reasons are:

(1) **Like energy and momentum, mass is not a constant of motion**. In classical physics, mass is regarded as an intrinsic property of an object, and thus, mass is not expected to change during motion. But in reality, we know this is not true. From Eq. (36), we see the moving mass of a particle is speed-dependent, i.e.,

$$m = \frac{m_0}{\sqrt{1 - v^2/c^2}}.$$

This can be rewritten as

$$\left(\frac{m_0}{m}\right)^2 = 1 - \frac{v^2}{c^2} \qquad (49)$$

(See Figure 7). It is apparent that the mass $m$ changes with the particle velocity $v$. This relationship has been confirmed in a number of experiments [25].

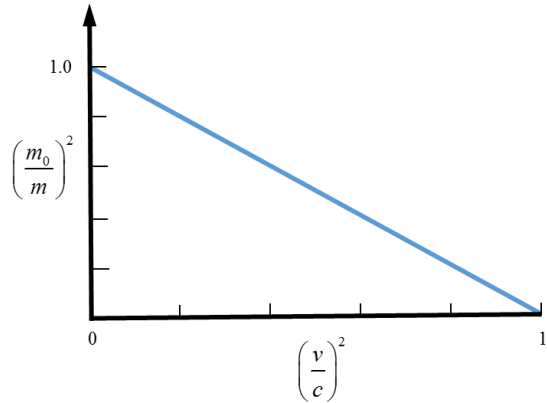

*Figure 7. Mass m can vary with velocity v. This is based on a plot of Eq. (49).*

(2) **The mass $m$ is not an independent parameter in the wave model**. In Section 3, we showed that mass is only a coefficient in the momentum versus velocity relation, i.e., $p = mv$. We know that $p$ is related to the wave vector $k$ of the particle and $v$ is defined as $v = d\omega/dk$, so mass is just a ratio between these two wave properties. Since $m$ is only a coefficient, there is no reason that it should be a constant (see Fig. 6).

(3) **Mass is just a measure of energy.** In Section 9 of this paper, we showed that the mass of a particle is closely related with the particle energy $E$, i.e., $E = mc^2$. So, $m$ is just equal to $E/c^2$. It is not an independent physical quantity; it is just a measure of energy. One may ask: Between $m$ and $E$, which one is more basic? If we accept that the particle is an excitation wave, then its energy is well defined; it is based on the Planck's relation: $E = \hbar\omega$. But one cannot say the same about the mass. *m* is not directly connected with the wave property other than through $E = mc^2$. Thus, *E* is more basic than *m* according to the wave model.

(4) **The rest mass of a particle is actually determined by the resting energy of that particle.** One may argue that although the mass *m* of a particle is not a constant, its rest mass $m_0$ is fixed with a particle. Could we regard the rest mass as an intrinsic property of an object? This is arguable, but it is just a semantic problem. As we have shown in Section 9, the particle is characterized by its resting energy $E_0$. Different kind of particles has different values of $E_0$. Since $m_0$ is just equal to $E_0/c^2$, one could say that different kind of particles have different values of $m_0$. But this statement is not more meaningful than our conclusion that the resting energy $E_0$ is the intrinsic property of a particle.

*11.2. Why no particle can travel faster than c?*

An advantage of the wave model is that it can naturally explain why no particle can travel faster than the speed of light. It is really based on three simple considerations:

(1) Since all particles are excitation waves of the vacuum, their travelling speeds are determined by the physical properties of the same vacuum medium. So, they must all have the same speed limit.
(2) Because the particle is a wave packet, the travelling speed is its "group velocity" instead of its "phase velocity". The group velocity is determined by $d\omega/dk$ and is generally slower than the phase velocity ($\omega/k$). The value of the phase velocity in our vacuum is known to be $c = 1/\sqrt{\mu_0 \varepsilon_0}$; it is the same for all particles (see Eq. (17)).
(3) For photon, its dispersion relation is $\omega = ck$. Its group velocity happens to be the same as its phase velocity. So it will always travel at the speed of *c*.

Thus, particles in general travel at a speed less than the speed of light (*c*); and, *c* is the ultimate speed for all particles. This conclusion was introduced in the STR as a postulate. But we can see here that it is actually a consequence of the fact that all particles are excitation waves of the vacuum.

*11.3. Why the mass of a particle is not a constant?*

Based on the above understanding, we can easily explain why the mass of a particle is speed-dependent. Several experiments have indicated that when a particle is accelerated to high speed, its mass appears to increase dramatically [25]. This phenomenon can be explained very easily from the wave view. Since all sub-atomic particles are excitation waves transmitting through the vacuum, their limiting speed of propagation is determined by the physical properties of the vacuum medium. This limiting speed is the speed of light *c*. When the particle speed *v* is much smaller than *c*, one can easily apply force to accelerate the particle. The energy received by the particle is used mainly to increase the particle speed. But this situation will change when the particle speed approaches that of light speed, because it will become very difficult to further increase *v*. That means, the particle will be very difficult to accelerate with the same amount of force. It thus appears that the inertial mass of the particle increases dramatically at high speed. In another word, the input energy is no longer used to drive the speed of the particle, instead, it appears that this additional energy is to increase the particle mass.

Therefore, the idea of regarding *m* as a constant is just an illusion in classical mechanics. If one regards the particle as a massive pointed object and *m* is its intrinsic property, one will not expect *m* to change. But if one understands that particle is just a wave, one will not have such an expectation.

### *11.4. Comparison with special relativity*

Historically, it was first pointed out by Einstein that the mass is not a constant of motion. He also suggested that the energy and mass could be related by the well-known formula: $E = mc^2$. In this work, we examined the relations between the energy, momentum and mass of a particle from the wave point of view. We found that the relations are very similar to those attributed to STR. However, there are important differences between our work and STR. *First, our relations are applicable only to a single particle. It does not apply to a bulk object. Second, our relations were derived based on the wave properties of the particle, not from relativity*. In our work, the main reason for a particle to behave differently from Newtonian mechanics is due to the fact that, in the microscopic level, a particle is a wave packet in nature; it does not behave as a mass point like a bullet.

Then, it will be helpful to design an experiment to test whether the non-Newtonian behavior of a particle is due to the relativistic effects or due to the wave effects. This could be done by determining whether there is a resting frame in our universe. An important postulate in STR is that there should be no resting frame. In this work, we assume that particles are excitation waves of the vacuum medium; such a medium will be very likely to form a resting frame. So, one can differentiate these two models by testing whether there is a resting frame in our universe or not.

One may ask that: Did the previous optical interferometer experiments, such as those conducted by Michaelson and Morley [26], already show that there is no resting frame in our universe? It is not true. As pointed out by Einstein in his original paper in 1905, such experiments only demonstrated that: "…*the same laws of electrodynamics and optics will be valid for all frames of reference…*" [27]. They did not demonstrate that all laws of physics are equivalent in every inertial frame. For a photon, it is understandable that its motion is frame-independent, since its rest mass is zero and its speed always equals to *c*. But for a particle with nonzero rest mass, the situation is different. Both of its *m* and *v* are speed-dependent. It is easy to see that *p* is different for different frames. Since $E^2 = p^2c^2 + m_0^2c^4$, it will mean that both *E* and *p* are dependent on the inertial frame chosen. As far as we know, it has not been demonstrated that the motion of a particle with nonzero rest mass is frame-independent.

A simple experiment to test whether there is a resting frame in our universe or not would be to measure the moving mass of a particle in a laboratory frame oriented in different direction in reference to the solar system. From Eq. (36), we know the moving mass of a particle is speed-dependent such that $m = \dfrac{m_0}{\sqrt{1 - v^2/c^2}}$. If STR is correct so that all inertial frames are equivalent, the moving mass should be the same regardless of the particle's moving direction. But if there is a resting frame in our universe, one will observe a difference in the moving mass when the traveling direction of the particle changes. This is because the value of $v$ in this case is a composite velocity of the particle velocity in respect to the laboratory frame plus the velocity of the laboratory frame itself.

We may add that, the question whether there is a resting frame or not is closely related with the question of whether our vacuum is totally empty. If one believes that the vacuum is a totally empty space, then there may not be a resting frame in our universe. But if one believes the so called "vacuum" in our universe has specific physical properties, it is natural to expect that such a vacuum can form a resting frame. In the Standard Model of cosmology today, the vacuum is definitely not empty. Many cosmology theories assume that the energy of our universe arises from the quantum fluctuation in the vacuum [28]; it cannot be regarded as an empty space. Also, according to the recent studies of CMB (cosmic microwave background), our universe is filled with visible matter, dark matter and dark energy [29,30]. Can they form a resting frame?

Furthermore, the vacuum is also not considered empty in the current theories of particle physics. For example, every oscillation mode in quantum electrodynamics should have a zero-point energy [31], which is supposed to be a part of the vacuum system. In the quantum field theory, the vacuum is always regarded as the ground state. The physical fields are just excitations above the vacuum [32]. Thus, the vacuum cannot be regarded as an empty space.

*11.5. Origin of dark matter according to the wave model*

In recent study of cosmology, it was discovered that our universe is not only composed of visible matters, but also dark matters and dark energy [29,30]. The name "dark matter" implies that it is invisible. This is because dark matter is transparent to electromagnetic radiation and thus cannot be detected using current imaging technologies. Its existence is indirectly inferred mainly from its gravitational effects on the universe's large-scale structure, such as the rotational motions of stars around galaxies and gravitational lensing [33,34]. The dark matter hypothesis plays an important role in current modelling of cosmic structure formation. It is also used to explain the anisotropies observed in the cosmic microwave background [29,30].

The standard model of cosmology indicates that the total mass/energy of the universe contains about 5% ordinary matter, 26% dark matter and 69% dark energy [29,30]. So the amount of dark matter is about five times of that of visible matter. At this point, very little is known about the dark matter. There are two major questions waiting to be answered: (1) What is dark matter composed of? Is it composed of particles? What could these particles be? (2) Why are there more dark matters than visible matters?

At present, the leading hypothesis about dark matter is that it is composed of weakly interacting massive particles (WIMPs) which interact only through gravity and the weak nuclear force [35]. These WIMPs are supposed to be new particles in the 100 GeV mass range. However, none of the experiments designed to detect WIMPs has produced any evidence for their existence [36-40]. Currently, there are several ongoing projects attempting to detect the WIMPs either directly or indirectly [41-43].

We think the matter wave model discussed in this work may provide a basis for searching new source of dark matters. According to our model, all particles in nature are excitation waves of the vacuum. Some of these excitation waves (particles) could behave as dark matter. These particles do not interact easily with other particles because they have certain properties. More specifically, these properties include:

- **Having no charge**. The dark matter (DM) particles have no electric charge. Thus, they do not interact with other charged particles through electromagnetic interactions. That means these DM particles will not interact with the electrons or nucleus of any atom. Also, because they have no electric charge, the DM particles will not interact with electro-magnetic radiation, including light. Thus, they will look dark.
- **Not being hadron**. These DM particles are unlikely to be hadrons. Since hadrons are made of quarks, any hadron can interact strongly with the atomic nuclei when it passes through an ordinary object. For example, neutron is a hadron without electric charge. It does not behave like dark matter because it can interact easily with the atomic nuclei of visible matters.
- **With small cross section**. For the dark matter not to interact with visible matter, its interaction cross section with particles of ordinary matters must be very small. That means, the DM particle may behave similar to neutrino, which has very small interaction cross section.
- **Possessing large resting energy**. Why the dark matter is dark? That is because we cannot see it. But what is *seeing*? It involves either absorption or scattering of the particle by the atoms.
  Take the photons for example. We see light because EM wave can interact with the orbital electrons of the atoms in our retina (or in a light sensor). For light to be absorbed by an orbital electron, it requires a matching of the particle energy ($E = \hbar\omega$) with the electron transition energy (the energy difference between the initial state and the final state of the electron, $\Delta E_{electron}$). In another word, absorption or emission of light requires $\hbar\omega \approx \Delta E_{electron}$ (see Figure 8). For most condensed matters, the energy level of an orbital electron (or electron bands) is in the order of several electron volts. If the energy of the incoming particle is much larger than the electron energy, the chance of absorption is extremely small. The particle will just pass by without interacting with the atom. This is why *X*-ray is more transparent to matter than visible light. If the DM particles have a large $E_0$, it will not interact easily with the orbital electrons of an atom or the band electrons of a solid. It thus would appear as "dark".

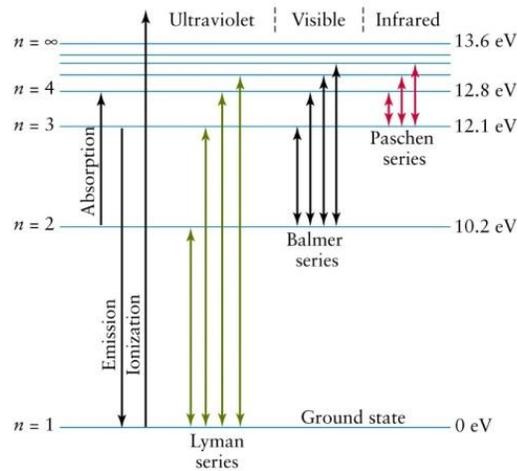

*Figure 8. Interaction between EM radiation and orbital electrons in hydrogen atoms. Examples showing that photons with different energies can interact with orbital electrons in different transition states. (Source: www.physast.uga.edu)*

The properties listed above are not unusual. It is reasonable to expect that many excitation waves in the vacuum can satisfy those requirements. Thus, the number of particles meeting these criteria could be very large. We did not notice them before because we cannot observe them. Also, since these DM particles have no charge, they cannot form atoms. And thus, they cannot aggregate together of form bulb objects.

Our model can also shed some light on the second question: *Why are there more dark matters than visible matters?* According to the wave model, all particles are excitation waves of the vacuum medium. Each particle represents a different excitation mode. Since there can be many excitation modes, the number of particle types in nature can be very large. From the discussion above, we suspect that many of these particle types can satisfy the requirements as dark matters. First, most of the excitation particles may not have electric charges. Many excitation waves can be photon-like or neutrino-like. Second, only some of the composite particles are hadrons; none of the simple particle is hadron. Thus, the number of non-hadron particles can be very large. Third, since particles are excitation waves, they normally do not interact with each other. Take the sound waves for example. Each sound can be transmitted independently without interference by other sounds. Imaging that when we talk with a friend in a noisy market place, there are many different sounds. Yet, we can still hear the voice of our friend. Hence, it is the norm that non-charged particles do not interact with each other. Finally, as shown in Figure 4, different particles can have different resting energies. Most of these resting energies are much higher than the binding energy of the orbital electrons in an atom. That means these particles are unlikely to be absorbed by the orbital electrons. Thus, they can pass through the matters easily without interacting with the atoms.

In summary, except for the charged particles, hadrons and photons, most particles do not normally interact with each other or interact with atomic matter. They will appear as dark matters in our experience, since they cannot be detected by our instruments. In another word, if we accept particles are excitation waves, the norm is that these waves do not interact. They are the majority. Only the special ones (i.e., charged particles, hadrons and photons) can interact with matter. They are the minority! *That is why there are more dark matters than visible matters*.

Hence, from this wave model, it is natural that most excitation waves do not interact with ordinary matters. They will be classified by us as "dark matter particles". The only way to detect these dark matter particles is through their gravitational effect. As we have discussed in Section 9 and 10, the gravitational effect is mainly due to energy-attracting-energy. Since like all excitation wave, these DM particles must possess energy, they will exhibit the gravitational effect. Thus, although these DM particles do not directly participate in the violent astronomical events such as star formation or supernova explosion (because they cannot form hadrons or atoms), they can affect the large scale structure development such as galaxy formation and galaxy clustering.

## 12. Conclusion

In this work, we proposed that all particles are excitation waves of the vacuum. In the microscopic view, a particle is just a wave packet. It behaves like a "corpuscular object" only in the macroscopic view. That is why a particle exhibits wave-particle duality. Because of its wave nature, the motion of a particle in the microscopic level will not follow the Newtonian mechanics. The major findings of this work include:

- In the wave view, "mass" can no longer be regarded as an intrinsic property of the particle. Instead, mass should be treated on the same footing as the energy and momentum of a particle.

- We showed that, *mass* is just a measure of *energy* in the wave view. In the classical limit, such an energy-related mass will appear in the equation of motion as an inertial mass.

- The wave representing a particle actually contains two types of energy: The "moving energy", which is directly related to the particle's momentum; and the "resting energy", which is independent of the momentum. We showed that, for a single particle, the *moving energy* and the *resting energy* form a two-dimensional Hilbert space.

- In this wave model, the *rest mass* of a particle ($m_0$) is associated with the *resting energy* of the particle ($E_0 = m_0 c^2$). Its "moving mass" ($m$) is related to the rest mass by the well-known formula, $m = m_0 / \sqrt{1 - v^2/c^2}$. In fact, many of the relativistic relations can be attributed to the fact that a particle is an excitation wave.

- The relations between *E*, *p* and *m* are the same for both *simple particles* (such as photons or electrons) and *composite particles* (such as hadrons or atomic nuclei).

- There is a smooth transition from wave mechanics to Newtonian mechanics. When the rest mass (or $E_0$) is small, the particle is *wave-like*; its motion can only be described by non-Newtonian mechanics. When the rest mass (or $E_0$) is very large, the particle is *corpuscle-like* (behaving like a massive pointed object). Its movement is described by Newtonian mechanics. Based on this consideration, photons and electrons normally do not behave like a classical particle. Objects composed of atoms, on the other hand, will behave like a classical object, because the resting energy of their nuclei are much larger than their moving energy (i.e., $E_0 \gg cp$). Thus, each atom behaves like a massive pointed object.

- We proposed that the "gravitational mass" is identical to the "energy-related mass". This infers that the source of gravitation is attributed to *energy-attracting-energy*, instead of *mass-attracting-mass*. This explains why light can bend in a gravitational field.

- This wave model offers a possible explanation to the origin of dark matters. It suggests that dark matters are composed of excitation waves in the vacuum that satisfy certain specific properties. It also gives an explanation why our universe has more dark matters than visible matters.

**Acknowledgements:** I am grateful to Prof. John A. Wheeler for his encouragement and comments during the early stage of this work. I thank Drs. H. E. Rorschach, Bambi Hu, Don Tow, Zhaoqing Zhang and Xiangrong Wang for their suggestions and comments. I also thank Ms. Lan Fu and Vivian Yu for assistance. A preliminary version of this work was presented in the 1984 Joint APS/AAPT Meeting [44]. This work was partially supported by the Research Grant Council of Hong Kong (*RGC 660207*) and the Macro-Science Program, Hong Kong University of Science and Technology (*DCC 00/01.SC01*).